\documentclass{article}
\usepackage{iccc}

\usepackage{times}
\usepackage{helvet}
\usepackage{courier}
\usepackage{url}
\usepackage{xspace}
\usepackage[english]{babel}

\usepackage[dvipsnames]{xcolor}
\usepackage{tikz}
\usetikzlibrary{calc}
\usetikzlibrary{positioning}
\usetikzlibrary{shapes.multipart}
\usetikzlibrary{fit,backgrounds}

\usepackage{multirow}
\usepackage{amssymb}
\usepackage[round,sort]{natbib}

\usepackage{amsmath,amsthm}
\newtheorem{definition}{Definition}

\usepackage[colorinlistoftodos]{todonotes}

\newcommand{\Amc}{\ensuremath{\mathcal{A}}\xspace}
\newcommand{\Bmc}{\ensuremath{\mathcal{B}}\xspace}
\newcommand{\Fmc}{\ensuremath{\mathcal{F}}\xspace}
\newcommand{\Imc}{\ensuremath{\mathcal{I}}\xspace}
\newcommand{\Qmc}{\ensuremath{\mathcal{Q}}\xspace}
\usepackage{onimage}

\title{An Argument-based Creative Assistant for Harmonic Blending \\
Paper type: System Description Paper
}
\author{
Maximos Kaliakatsos-Papakostas$^a$, Roberto Confalonieri$^b$, Joseph Corneli$^c$,\\
\Large{\textbf{Asterios Zacharakis$^a$ and Emilios Cambouropoulos$^a$}}\\
$^a$Department of Music, Aristotle University of Thessaloniki, Greece\\
$^b$Artificial Intelligence Research Institute, IIIA-CSIC, Bellaterra (Barcelona), Spain\\
$^c$Department of Computing, Goldsmiths, University of London, UK\\
$^a$\small{\texttt{\{max,aszachar,emilios\}@mus.auth.gr}}\\
$^b$\small{\texttt{confalonieri@iiia.csic.es}}\\
$^c$\small{\texttt{j.corneli@gold.ac.uk}}
}
\begin{document} 
\maketitle
\begin{abstract}
Conceptual blending is a powerful tool for computational creativity where, for example, the properties of two harmonic spaces may be combined in a consistent manner to produce a novel harmonic space. However, deciding about the importance of property features in the input spaces and evaluating the results of conceptual blending is a nontrivial task. In the specific case of musical harmony, defining the salient features of chord transitions and evaluating invented harmonic spaces requires deep musicological background knowledge. In this paper, we propose a creative tool that helps musicologists to evaluate and to enhance harmonic innovation. This tool allows a music expert to specify arguments over given transition properties. These arguments are then considered by the system when defining combinations of features in an idiom-blending process.  A music expert can assess whether the new harmonic idiom makes musicological sense and re-adjust the arguments (selection of features) to explore alternative blends that can potentially produce better harmonic spaces. We conclude with a discussion of future work that would further automate the harmonisation process.
\end{abstract}


\section{Introduction}\label{sec:introduction}

The invention of new harmonic spaces in this paper is conceived as a computational creative process according to which a new harmonic idiom is created by means of blending the `atoms' of harmony, i.e.,\ transitions between chords. The blended transitions are created by combining the features characterising pairs of transitions belonging to two idioms (expressed as sets of potentially learned transitions) according to an amalgam-based algorithm~\citep{eppe_asp_amalgamation_2015,confalonieriIJCAI15} that implements~\cite{FaTu02}'s theory of conceptual blending. The transitions are then used in an extended harmonic space that accommodates the two initial harmonic spaces, linked with the new blended transitions.

When modeling creative processes computationally, one of the key questions is how good are the created artefacts. The approach to evaluation that has been applied most frequently within computational creativity requires a human to evaluate attributes of the created work or the system's operation. Basic measures consider the \emph{typicality} of a generated artefact within a particular genre, or the \emph{quality} of the generated work according to the users' aesthetic judgement \citep{ritchie07}. 

In music blending, the evaluation of artefacts is not a trivial matter. This is due not only to the time evolving nature of the final output, but also to the lack of clearly defined criteria for their assessment. In the particular case of transition blending, the evaluation of the blends is of key importance, in order to produce musically meaningful extended harmonic spaces. To evaluate the set of blended transitions and the corresponding generated extended harmonic space, several musical features need to be taken into account according to indications by musicologists. The importance of each particular feature, however, is not known in advance and musicologists need to make adjustments by experimenting with a large set of test cases. 

\begin{figure}[!t]
\centering
\begin{tikzonimage}[width=.40\textwidth]{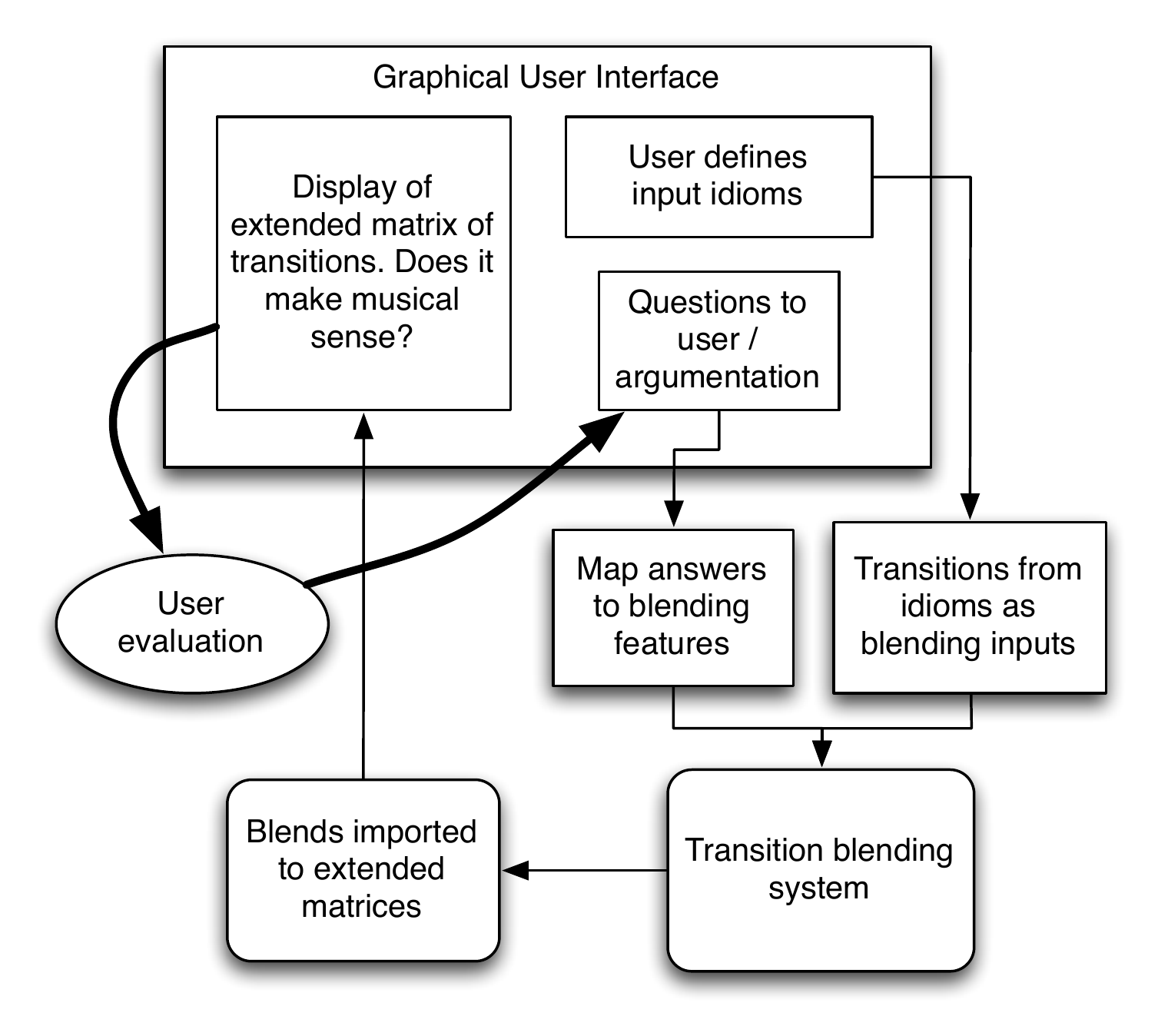}[every node/.style={text=gray,font=\footnotesize}
]
    \node at (0.78,0.93) {$\alpha$};
    \node at (0.75,0.675) {$\beta$};
    \node at (0.73,0.86) {$\gamma$};
    \node at (0.815,0.098) {$\epsilon$};
    \node at (0.4,0.115) {$\zeta$};
    \node at (0.244,0.415) {$\iota$};
\end{tikzonimage}
\caption{A schematic diagram of the system's workflow.}
\label{fig:systemOverview}
\end{figure}
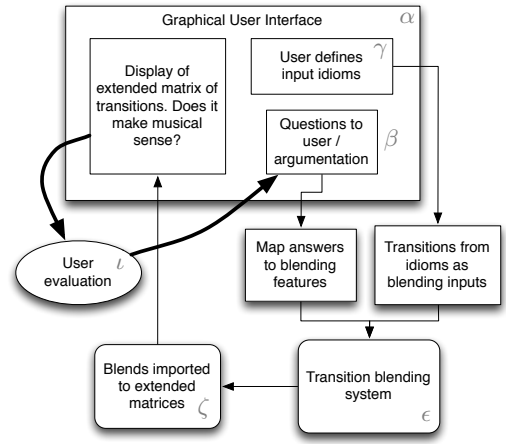

To ease this task, in this paper, we propose a creative tool (Figure~\ref{fig:systemOverview}) that assists a musicologist with the evaluation of harmonic blends. The system allows a musicologist  to specify {\em arguments} --- abstracting the properties of chords and transitions --- and to use them for iterative evaluation of the blended outcome, based on the transitions that the system proposes in order to connect two (potentially remote) harmonic spaces.

Using arguments to make and explain decisions has been proposed and explored in Artificial Intelligence \citep{CapDun07}, where an argument is a reason for believing a statement, choosing an option, or doing an action. In most existing works on argumentation, an argument is either considered as an abstract entity whose origin and structure are not defined \citep{Dung95}, or it is a logical proof for a statement where the proof is built from a knowledge base \citep{AmgPra09}.  


In our approach, arguments encapsulate desirable properties that the user would like to have in the resulting transition blends. Arguments are specified by the user by `answering'  specific questions over the features of the idioms selected as input for the transition blending process. Providing some higher level arguments as inputs to the system is equivalent to allowing a musical expert to interact with it in a language he/she understands. This offers the user a flexible way to adjust the harmonic blending properties according to different input scenarios in order to improve the creativity of the system. Extended experimentation with the system ---by making use of the available arguments--- can enable music experts to provide valuable feedback regarding the functionality of transition features, thus directly intervening in the blending process by answering simple questions. In a future scenario, the assessment of the system will be based on merely musicological criteria that should be more clearly defined.

The paper is organised as follow. In the next section, we describe the harmonic blending creative process embedded in a creative assistant tool we implemented. Next, we describe the methodology of transition blending and extending harmonic spaces. We show how user arguments are used to evaluate transition blends based on two criteria resembling two of the optimality principles of conceptual blending. Then, we present a process-based system evaluation that focuses on the creative acts of programmers \citep{colton-assessingprogress}. This evaluation is helpful in guiding further developments of the system. These are discussed in a concluding section.


\section{System overview and test cases}\label{sec:system}

Figure~\ref{fig:systemOverview} illustrates a diagram of the presented system. The user (music expert) interacts with the system through the Graphical User Interface (GUI), where she/he selects two initial idioms (harmonic spaces) in $\gamma$ and defines the important features used in conceptual blending by answering to specific questions (argumentation) in $\beta$. The selected initial idioms are described as sets of chord transitions, while the provided answers to questions are mapped to enabling/disabling features of transitions (see Section `Chord transitions description and blending') that define the outcome of transition blending (see Section `Evaluation of transition blending via arguments').

Afterwards, pairs of transition in the two initial harmonic spaces are given as inputs to the transition blending system in $\epsilon$ where new transitions are invented through conceptual blending. These transitions are then integrated into an \textit{extended} musical idiom that includes the initial idioms selected by the user, while the role of the new transitions is to provide musically meaningful connections between the initial harmonic spaces. The created extended idiom is displayed to the user in the GUI in terms of a transition matrix (see Section `From transition blends to transition matrices'). By observing the matrix, the music expert evaluates ($\iota$) the results produced by the current blending setup, i.e.,\ the given questions to the argumentation module ($\beta$), and re-adjusts her/his answers in $\beta$ accordingly.

Several scenarios for initial idiom combinations are available to the user. The system included several harmonic blending test cases according to which the user could blend simple `artificial' harmonic spaces as well as harmonic spaces trained from data in different tonalities. The artificial harmonic spaces are constructed to include simple transitions in order to allow clear interpretations of the results, e.g.,\ a C major space included the chords C, F and G7. Among the trained idioms that have been examined, there are sets of Bach chorales in major and minor mode, and sets of modal chorales in several modes.

The test cases, in which harmonic spaces in different tonalities are blended, resemble the musical task of finding transition paths for tonality modulations (changing the tonality of a given harmonic space). This task allowed music experts to identify arguments for defining the features of transition blending that connect potentially remote harmonic spaces (e.g.,\ C major with F$\sharp$ major) in a manner that is explainable in music theory in terms of tonality modulations. Through the processes offered by the system, the music experts were able to come to conclusions about what transition features are important for constructing meaningful connections between different combinations of pairs of initial harmonic spaces.

\section{Methodological aspects of transition blending and extending harmonic spaces}\label{sec:approach}


The cognitive theory of conceptual blending by \cite{FaTu02} has been extensively used in linguistics, music composition~\citep{zbikowski_conceptualizing_2005}, music cognition~\citep{antovic_musical_2009,antovic2011musical} and other domains mainly as an analytical tool, which is useful for explaining the cognitive process that humans undergo when engaged in creative acts. According to this theory, human creativity is modeled as a process by which a new concept is constructed by taking the commonalities among two {\em input spaces} into account, to form a so-called {\em generic space}, and by projecting their non-common structure in a selective way to a novel blended space, called a {\em blend}. 

\begin{figure}[!t]
\includegraphics[width=0.40\textwidth]{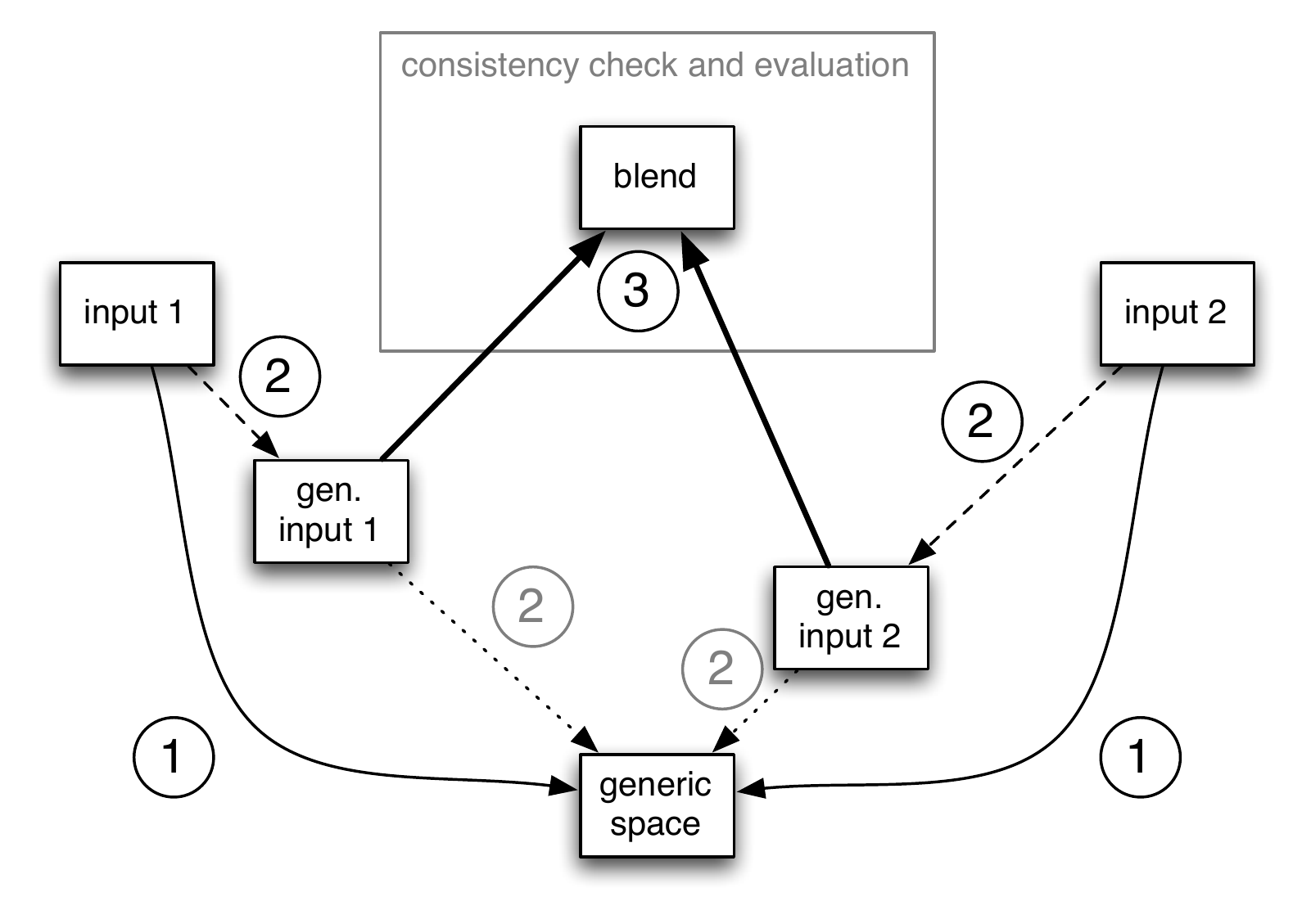}
    \caption{Conceptual blending based on amalgam. The generic space is computed (1) and the input spaces are successively generalised (2), while new blends are constantly created (3). Some blends might be inconsistent or purely evaluated according to blending optimality principles or domain specific criteria.}
    \label{fig:blendingGeneral}
\end{figure}

In computational creativity, conceptual blending has been modeled by \cite{goguen06a} as a generative mechanism, according to which input spaces are modeled as \textit{algebraic specifications} and a blend is computed as a categorical \textit{colimit}. A computational framework that extends Goguen's approach has been developed in the context of the COncept INVENtion Theory\footnote{\url{http://www.coinvent-project.eu}} (COINVENT) project~\citep{coinvent14} based on the notion of {\em amalgams}~\citep{OnPl10}. According to this framework, \textit{input spaces} are described as sets of features, properties and relations, and 
an \textit{amalgam}-based workflow finds the blends~\citep{eppe_asp_amalgamation_2015,confalonieriIJCAI15}. The amalgam-based workflow generalises input concepts until a generic space is found and `combines' generalised versions of the input spaces to create blends that are consistent or satisfy certain properties that relate to the knowledge domain (Figure~\ref{fig:blendingGeneral}).\footnote{In the process of blending through amalgams, the notions of `amalgam' and `blend' are the same. Therefore, in the following paragraphs they are used interchangeably.} 

\begin{figure}[!t]
\centering
	\includegraphics[width=.45\textwidth]{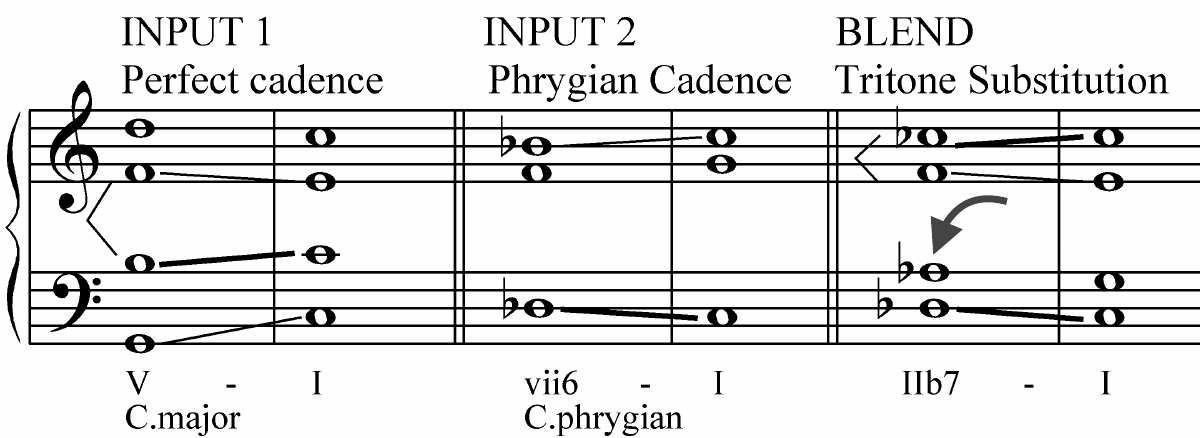}
\caption{Example of blending cadences, which are special case of transitions, where blending the perfect and the Phrygian produce the tritone substitution cadence blend.}
\label{fig:cadsExample}
\end{figure}

Amalgam-based conceptual blending has been applied to invent chord cadences \citep{chordBlendingIJCAI15,cads_ismir_2015}.  In this setting, cadences are considered as special cases of chord transitions---pairs of chords, where the first chord is followed by the second one--- that are described by means of features such as the roots or types of the involved chords, or intervals between voice motions, among others. When blending two transitions, the amalgam-based algorithm first finds a generic space between them (point 1 in Figure~\ref{fig:blendingGeneral}). For instance, in the case of blending the perfect with the Phrygian cadences (Figure~\ref{fig:cadsExample})---described by the transitions I$_1$: G7 $\rightarrow$ C and I$_2$: B$\flat$m $\rightarrow$ C5 respectively--- their generic space consists of any transition that has a second chord with the root note C, since this is the root note of both inputs' second chords (C and C5).

After a generic space is found, the amalgam-based process computes the amalgam of two input spaces by \textit{unifying} their content. If the resulting amalgam is inconsistent, then it iteratively generalises the properties of the inputs (point 2 in Figure~\ref{fig:blendingGeneral}), until the resulting unification is consistent (point 3 in Figure~\ref{fig:blendingGeneral}). For instance, trying to directly unify the transitions I$_1$: G7 $\rightarrow$ C and I$_2$: B$\flat$m $\rightarrow$ C5 would yield an inconsistent amalgam, since a transition cannot both include and \textit{not} include a leading note to the second chord's tonic (which is a property of I$_1$ and the I$_2$ respectively). Therefore, the amalgam-based process generalises the clashing property in one of the inputs (e.g., the property describing the absence of leading note would be left empty in I$_2$) and tries to unify the generalised versions of the inputs again. After a number of generalisation steps are applied (point 2 in Figure~\ref{fig:blendingGeneral}), the resulting blend is consistent (point 3 in Figure~\ref{fig:blendingGeneral}). However, it may be the case that the blend is not complete, in the sense that this process may have generated an over-generalised term. 

\textit{Blending completion}~\citep{FaTu02} is a domain-specific process that uses background knowledge to consistently assign specific properties to generalised terms. In the hitherto discussed example, blend completion is used for completing the $A\flat$ note as the chord's fifth in blending the perfect and Phrygian cadence in order to obtain the tritone substitution cadence (Figure~\ref{fig:cadsExample}).

After several blends have been computed, an evaluation process ranks them according to some optimality principles~\citep[Chapter 16]{FaTu02}. These principles are a necessary aspect of conceptual blending since they allow to filter interesting blends from the (potentially too) many possible ones
\footnote{The amalgam-based algorithm produces many blends by following alternative generalisation paths. 
}. A complete description of optimality principles is out of the scope of this paper and the reader is referred to~\cite{goguenharrell10} for applications of such principles in the \textit{Alloy} algorithm. We give, however, two extreme examples of `bad blends' for clarifying the importance of using optimality principles in conceptual blending. 
\begin{itemize}
	\item \textit{Example 1, violating the symmetry principle:} Each of the input spaces is a trivial form of a blend. This is a bad blend, because no information from the other input spaces is considered. 
	\item \textit{Example 2, violating the web principle:} Consider a blend that includes all properties of the generic space, but all other information are filled by properties that are not included in any of the input spaces. This blend has the least possible connections with the input spaces and therefore, the least amount of information from the inputs is identifiable in this blend.
\end{itemize}
%
These examples suggest two criteria for ranking the blends; we provide a computational characterisation of them below.

\subsection{Chord transitions description and blending}\label{subsec:transBlending}

Individual chord transitions are the `atoms' of the methodology followed herein to construct new transition matrices. Specifically, transition sets from two musical idioms provide input transitions for blending, producing a list of blended transitions that are afterwards embeded in an extended harmonic space. This methodology is described briefly in the next section 
while some definitions regarding chord transitions follow.

\begin{definition} A chord transition $c$ is described by a set of features \Fmc.
\end{definition}
 
\noindent In this work a transition is represented by $17$ features. Features $1$-$6$ refer to the involved chords. Features $8$ to $10$ indicate changes during the transitions and are based on the Directed Interval Class (DIC) vector~\citep{cambouropoulosDIC_12, cambouropoulosDIC_13}. Feature $7$ accounts for the change that occurred regarding the chords' root notes. The features considered important in this work are the following:
\begin{enumerate}
	\item \textit{fromRoot}: the root pitch class of the first chord,
	\item \textit{toRoot}: the root pitch class of the second chord,
	\item \textit{fromType}: the type of the first chord (GCT base),
	\item \textit{toType}: the type of the second chord (GCT base),
	\item \textit{fromPCs}: the pitch classes included in the first chord,
	\item \textit{toPCs}: the pitch classes included in the second chord,
	\item \textit{DICinfo}: the DIC vector of the transition,
	\item \textit{DIChas0}: Boolean value indicating whether the DIC of the transition has $0$,
	\item \textit{DIChas1}: Boolean value indicating whether the DIC of the transition has $1$,
	\item \textit{DIChasMinus1}: Boolean value indicating whether the DIC of the transition has $-1$,
	\item \textit{ascSemZero}: Boolean value indicating whether the first chord has the relative pitch class value 11,
	\item \textit{descSemZero}: Boolean value indicating whether the first chord has the relative pitch class value 1,
	\item \textit{semZero}: Boolean value indicating whether the first chord has the relative pitch class value 11 or 1,
	\item \textit{ascSemNextRoot}: Boolean value indicating whether the first chord has a pitch class with ascending semitone relation with the pitch class of the second chord's root,
	\item \textit{descSemNextRoot}: Boolean value indicating whether the first chord has a pitch class with descending semitone relation with the pitch class of the second chord's root,
	\item \textit{semNextRoot}: Boolean value indicating whether the first chord has a pitch class with ascending or descending semitone relation with the pitch class of the second chord's root, and
    \item \textit{5thRootRelation}: Boolean value indicating whether the first chord's root note is a fifth above of the second's.
\end{enumerate}

Each feature can be considered as a function that assigns a value to a chord transition $c$. Features' values are defined differently depending on the properties they represent. For instance, features $3$ to $8$ are set-value functions that assign a set of values to a chord. We refer to them as $F_i(c)$. The value of the feature $7$ is a vector and we refer to it as $\vec{f}(c)$. Finally, all the other features are binary functions and we refer to them as $f_i(c)$.

\subsection{From transition blends to transition matrices}\label{subsec:toMatrices}


In the literature, an effective and common way to describe chord progressions in a music idiom in a statistical manner is by using Markov models (see \citet{simon2008mySong,kaliakatsos_cHMM_14}, among others). Such models reflect the probabilities of each chord following other chords in the idiom, as trained or statistically measured throughout all the pieces in the examined idiom. First-order Markov models, specifically, indicate the probability of transitions from one chord to another, disregarding information about previous chords. Therefore, individual transitions play an important role on indicating particular characteristics of an idiom.

A convenient way to represent a first order Markov model is through transition matrices, which include one respective row and column for each chord in the examined idiom. The probability value in the $i$-th row and the $j$-th column exhibits the probability of the $i$-th chord going to the $j$-th ---the probabilities of each row sum to unit. The utilised chords are actually represented by chord group exemplars, obtained by the method described in~\cite{gctEval_ismir_2015}, while transitions between chords that pertain to the same chord group are disregarded (this neutralises the diagonal). The representation of chords is based on the General Chord Type representation~\citep{cambouropoulosGCT_14}.


Then, an important question is: {\em How would a blended idiom be expressed in terms of a transition matrix, provided that the transition matrices of two initial idioms are available?} 

\begin{figure}[!t]
\centering
	\includegraphics[width=.40\textwidth]{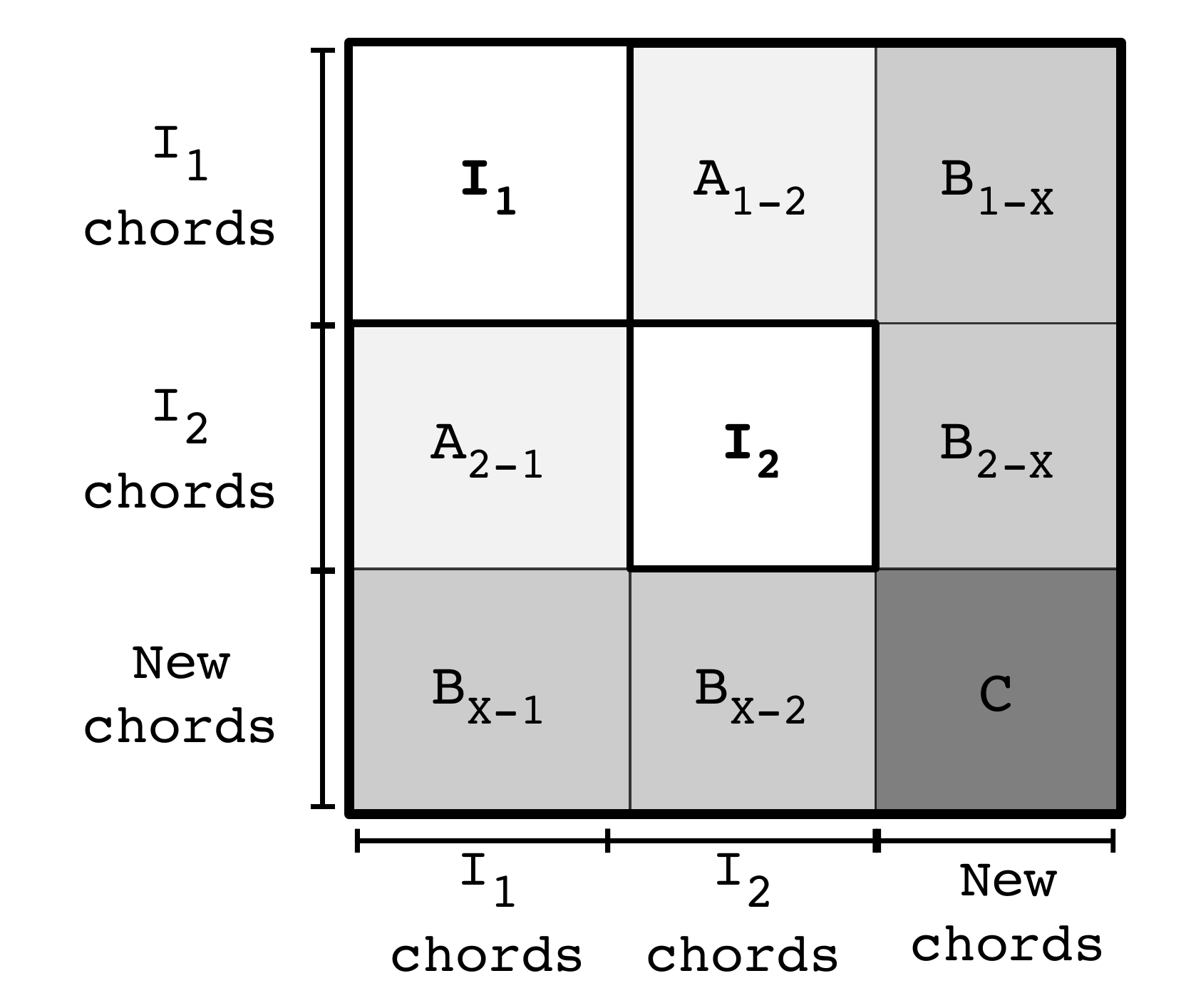}
\caption{Graphical description of an \textit{extended} matrix that includes transition probabilities of both initial idioms and of several new transitions generated through transition blending. These new transitions allow moving across the initial idioms, creating a new extended idiom that is a superset of the initial ones.}
\label{fig:extendedMat}
\end{figure}

Among many possible answers, the idea examined in the present system is to create an \textit{extended} transition matrix that includes not only an altered version of the initial ones, but also new transitions that allow moving across chords of the initial idioms by potentially using new chords. The examined methodology uses transition blending to create new transitions that: (a) maximally preserve the common parts of transitions between the two initial spaces, and (b) incorporate blended characteristics for creating a smooth `morphing' harmonic effect when moving from chords of one space to chords of the other. An abstract illustration of an extended matrix is given in Figure~\ref{fig:extendedMat}.

By analysing the graphical representation of an extended matrix as depicted in Figure~\ref{fig:extendedMat} the following facts are highlighted:
\begin{enumerate}
	\item By using transitions in $\mathtt{I_i}$, only chords of the $i$-th idiom are used. When using the transition probabilities in $\mathtt{I_i}$, the resulting harmonisations preserve the character of idiom $i$.
	\item Transitions in $\mathtt{A_{i-j}}$ create direct jumps from chords of the $i$-th to chords of the $j$-th idiom. If a blended transition happens to be in $\mathtt{A_{i-j}}$ there is no need for further considerations -- such a transition can be included in the extended matrix.
	\item Transitions in $\mathtt{B_{i-X}}$ constitute harmonic motions from a chord of idiom $i$ to a newly created chord by blending. Similarly, transitions in the $\mathtt{B_{X-j}}$ arrive at chords in idiom $j$ from new chords. For moving from idiom $i$ to idiom $j$ using one external chord $c_x$ that was produced by blending, a chain of two transitions is needed: $c_i$ $\rightarrow$ $c_x$ followed by a transition $c_x$ $\rightarrow$ $c_j$, where $c_i$ in idiom $i$ and $c_j$ in idiom $j$ respectively. A chain of two consecutive transitions with one intermediate external chord from chords of $i$ to chords of $j$ will be denoted as $\mathtt{B_{i-X-j}}$.
	\item Sector $\mathtt{C}$ transitions incorporate pairs of chords that exist outside the $i$-th and $j$-th idioms. Having two external chords, transitions in $\mathtt{C}$ violate our hypothesis for moving from one known chord sets to the other using one new chord at most; therefore, blends resulting to $\mathtt{C}$-type blends are disregarded.
\end{enumerate}

Based on this analysis of the extended matrix, a methodology is proposed for using blends between transitions in $\mathtt{I_1}$ and $\mathtt{I_2}$. Thereby, transitions in $\mathtt{I_1}$ are blended with ones in $\mathtt{I_2}$ and a number of the best blends is stored for further investigation, creating a pool of best blends. Based on multiple simulations, a large number of the best blends (i.e.\ 100) in each blending simulation should be inserted in the pool of best blends (\Bmc), so that several commuting scenarios can be created between the initial transition spaces. Thus, a greater number of blends in the pool of best blends introduces a larger number of possible commuting paths in $\mathtt{A_{i-j}}$ or in $\mathtt{B_{i-X-j}}$.

\subsection{Evaluation of transition blending via arguments}\label{subsec:argumentation}

By applying the aforementioned blending process a pool of best blends is created that is afterwards used for connecting the transition blocks of two initial idioms, through forming an extended matrix. When a music expert is using the system, she/he is able to select pairs of initial input idioms, choose which aspects of blending are important through arguments (analysed in the following paragraphs) and evaluate/re-adjust this choice by observing the produced results in the extended matrix. 



The user evaluates the importance of several transition features by answering questions based on the connecting transitions produced by blending in the extended matrix. The features related to the transitions and their constituent chords are classified into 9 questions~(Table~\ref{tbl:abstraction}). 
\begin{description}
	\item[Q1:] Are roots and types of chords important?
	\item[Q2:] Are individual pitch classes of chords important?
	\item[Q3:] Are repeating pitch classes in transitions important?
	\item[Q4:] Are semitone steps in transitions important?
	\item[Q5:] Are tone steps in transitions important?
	\item[Q6:] Are the intervalic contents of transitions important?
	\item[Q7:] Are semitone motions to the tonic important?
	\item[Q8:] Are semitones to the second chord's root important?
	\item[Q9:] Are motions of the chord roots by $5^\text{th}$ important?
\end{description}
The first two questions concern characteristics of the chords that constitute the transition, mapping the user answers to features from 1 to 6, while the remaining seven questions concern intervalic changes that occur within the transition, mapping the user answers to features from 7 to 17. Relating questions to transition features was performed with the involvement of music experts, to ensure that the mapping is as accurate and as informative to the user as possible.

\begin{table}[!t]
\begin{tabular}{lll}
\hline \hline
Question & Chord Properties & Transition Changes \\ \hline
\multirow{4}{*}{Q1} & $\textit{fromRoot}$\\
    			      & $\textit{toRoot}$ \\
			      & $\textit{fromType}$\\
			      & $\textit{toType}$ & $$ \\ \hline
\multirow{2}{*}{Q2} & $\textit{fromRelPCs}$\\
    			      & $\textit{toRelPCs}$ & $$ \\ \hline
Q3 & $$ & $\textit{DIChas0}$  \\ \hline
Q4 & \multirow{2}{*}{} & $\textit{DIChas1}$ \\
				  & $$ & $\textit{DIChasN1}$\\ \hline
Q5 & \multirow{2}{*}{} & $\textit{DIChas2}$ \\
				  & $$ & $\textit{DIChasN2}$\\ \hline
Q6 & $$ & $\textit{DICinfo}$  \\ \hline
Q7 & \multirow{3}{*}{} & $\textit{ascSemZero}$ \\
				  & $$ & $\textit{descSemZero}$\\ 
				  & $$ & $\textit{semZero}$\\ \hline
Q8 & \multirow{3}{*}{} & $\textit{ascSemNextRoot}$ \\
				  & $$ & $\textit{descSemNextRoot}$\\ 
				  & $$ & $\textit{semNextRoot}$\\ \hline
Q9 & $$ & $\textit{5thRootRel}$  \\ \hline
\end{tabular}
\caption{Abstraction of chords' and transition changes' features.}
\label{tbl:abstraction}
\end{table}

We denote the set of questions available to the user as \Qmc. When a user selects a question, an argument is automatically generated. For the sake of this paper, an argument is defined as follows.

\begin{definition}\label{def:argument}
An argument $A$ is a tuple $\langle q, F \rangle$, where $q \in \Qmc$ and $F \subset \Fmc$. 
\end{definition}

\noindent The user can specify at most $9$ arguments, each of them is mapped to a set of properties. The set of user arguments $\{A_1,\ldots,A_9\}$ corresponding to answers to $\Qmc$ will be denoted by \Amc.  We assume to have a function $\mathsf{\psi}: \Amc \rightarrow \Fmc$ that returns the set of chord and transition properties associated with an argument (e.g, for the purposes of the current analysis, Table \ref{tbl:abstraction} specifies $\psi$ as a look-up function). The arguments are used to compute two criteria in order to rate a blend: \textit{total association} and  \textit{symmetry}. 

The total association indicates the total number of properties that a blend inherits from the inputs. A blend with higher input associations is preferable since it is structurally more deeply related with the inputs. The total association is calculated by taking the individual association of a blend w.r.t. the input chord transitions into account. The individual association of a blend $b$ w.r.t. to an input $I$, denoted as $\mathsf{a}(b,I)$, is defined as:
\begin{equation*}\label{eq:inputassoc}
	\mathsf{a}(b,I) = \sum_{A_i \in \Amc}\mathsf{Val}(A_i,b,I)
\end{equation*}
\noindent where $\mathsf{Val}: \Amc \rightarrow \mathbb{R}$ is a function that takes an argument as input and aggregates the values of the chord and transition change properties associated with the argument, by interpreting them according to some music background knowledge. Depending on the type of argument, $\mathsf{Val}$ is defined in different ways. 

When an argument refers to the roots and types of chords ($A_1$), $\mathsf{Val}$ is defined as: 
\[
\mathsf{Val}(A_1,b,I) =   \sum_{F_j \in \psi(A_1)}{  \mathsf{equals}(F_j(I), F_j(b)) }
\]

\noindent The value of $A_1$ is calculated by counting how many properties ---among $\textit{fromRoot}$, $\textit{toRoot}$, $\textit{fromType}$ and $\textit{toType}$--- are equals between a blend $b$ and an input $I$.  $\mathsf{equals}$ is a function that returns $1$ when two sets are equals and $0$ otherwise.

When an argument refers to the individual pitch classes of chords ($A_2$),  $\mathsf{Val}$ is defined as:
\[
\mathsf{Val}(A_2,b,I) =   \sum_{F_j \in \psi(A_2)}{ \left | F_j(I) \cap F_j(b) \right | } 
\]

\noindent The value of $A_2$ is calculated as the number of elements that are common in the set-value properties $\textit{fromRelPCs}$ and $\textit{toRelPCs}$ of a blend $b$ and an input $I$.

When an argument refers to the intervalic contents of transitions ($A_6$),  $\mathsf{Val}$ is defined as: 
\[
\mathsf{Val}(A_6,b,I) =  \mathsf{norm}_{[0,1]}(\rho_{\vec{f}(I),\vec{f}(b)})
\]

\noindent The value of $A_6$ is calculated as the Pearson's correlation coefficient of the vector-value property $\textit{DICinfo}$ of a blend $b$ and an input $I$. Higher correlations in the DIC vectors of two transitions indicate higher resemblance; $\mathsf{norm}$ is a function that normalises the Pearson's coefficient from the interval $[-1,1]$ to the interval $[0,1]$.

\noindent For all the other types of arguments, $\mathsf{Val}$ is defined as:
\[
\mathsf{Val}(A_i,b,I) =  \sum_{f_j\in \psi(A_i)}{ 1 -  (f_j(I) - f_j(b)) } 
\]


\noindent Based on the above definitions, the {\em total association} value is 
the sum of the individual associations.

\begin{equation*}\label{eq:inputassoc}
	\mathsf{assoc}(b) = \sum_{I_i \in \Imc}\mathsf{a}(b,I_i)
\end{equation*}

\noindent where \Imc is the set of input spaces, containing in this specific case, $I_1$ and $I_2$.

Symmetry, on the other hand, reflects the balance of properties that a blend inherits from both input spaces. A blend has a high symmetry when it inherits an almost equal proportion of properties from both input spaces. 
Blends having higher symmetry are preferred to those with lower symmetry, since a high symmetry reflects a stronger hybridisation of structural characteristics. Hybridisation is an important principle to evaluate transition blends. 

The blend symmetry is defined 
in terms of  its `asymmetry'. The asymmetry of a blend w.r.t. the inputs, denoted as $\mathsf{asym}(b)$, 
is calculated as:
%
\begin{equation*}\label{eq:asymmetry}
	\left| \frac{\mathsf{a}(b,I_1)^2 + \mathsf{a}(b,I_1) \mathsf{a}(b,I_2)}{\mathsf{a}(b,I_1)^2+\mathsf{a}(b,I_2)} - \frac{\mathsf{a}(b,I_2)^2+\mathsf{a}(b,I_1)\mathsf{a}(b,I_2)}{\mathsf{a}(b,I_2)^2+\mathsf{a}(b,I_1)} \right| 
\end{equation*}
\noindent The value of $\mathsf{asym}(b)$ is defined in $[0,1]$, where $0$ stands for a perfect symmetry (equal association with both inputs) and $1$ stands for total asymmetry (association only with one input). Additionally, the non-absolute version of the above equation suggests the prevailing input, with a negative value indicating dominating association of the blend with the first input and a positive value contrarily.

The total rate of a blend is computed by taking the input association and asymmetry values into account.
\begin{equation*}\label{eq:rating}
	\mathsf{rate}(b) = \frac{\mathsf{assoc(b)}  (1-\mathsf{asym}(b))}{\mathsf{assoc(b)} + (1-\mathsf{asym}(b))} 
\end{equation*}
The above expression 
promotes pairs of association and symmetry that are both high, while a simple sum would allow a low value of the one to be covered by the other. 

Finally, a decision making criterion to compare any pair of blends $b_1$, $b_2 \in \Bmc$ can be defined as follows.

\begin{definition}[Decision criterion]
A blend $b_1$ is preferred to a blend $b_2$ if and only if $\mathsf{rate}(b_1)  \geq \mathsf{rate}(b_2)$. 
\end{definition} 

It is worthy to notice that the above criterion guarantees that any pair of blends is comparable, and, consequently, it allows to decide which blends are the best ones. This is an important property for blend evaluation and, generally, for approaches to argumentation-based decision making \citep{AmgPra09,geffner}.





\section{System evaluation}\label{sec:evaluation}

Referring to Figure \ref{fig:systemOverview}, via the interface $\alpha$, the user has access to modules $\gamma$, and $\beta$ which can be used to specify \emph{concepts} that will inform the resulting product, namely, the input idioms and arguments that impose constraints on the generated blend.  These are translated by the system into process-friendly formats.
Module $\epsilon$ embodies the (process-level) concept of a system that make use of the supplied idioms and the blending properties to generate \emph{example} transition matrices,  $\zeta$.
In the current version of the system, these transitions are evaluated by the user (music expert) in step $\iota$ using sophisticated harmonic knowledge that reflects historically established musical \emph{aesthetic}. The user can then return to the GUI $\alpha$, and adjust the settings of $\gamma$ and $\beta$ to regenerate the transitions.

\begin{figure}
\centering
\begin{minipage}{1\columnwidth}


\resizebox{1\columnwidth}{!}{%
\begin{tikzpicture}[
single/.style={draw, anchor=text, rectangle},
double/.style={draw, anchor=text, rectangle split,rectangle split parts=2},
triple/.style={draw, anchor=text, rectangle split,rectangle split parts=3},
quadruple/.style={draw, anchor=text, rectangle split,rectangle split parts=4}
]
\node[single,scale=0.3] (first) at (0, 0) {
  \tikz{
\node[triple] (firstA) at (-2,0) {$A^{\beta,\gamma}_g = \langle T(\overline{C^{\beta}_g},\overline{C^{\gamma}_g})\rangle$
  \nodepart{second}{$\langle E^{\epsilon}_g\rangle^* = \overline{C^{\epsilon}_p}(a^{\beta,\gamma}_g)$}
  \nodepart{third}{$e^{\zeta}_g = \langle S[a^{\beta,\gamma}_g]({e^{\epsilon}}^*) \rangle$}
};
\node[single,right=6mm of firstA.three east] (firstB) {$\langle \overline{A^{\iota}_g}(e^{\zeta}_g)\rangle$
};
\draw [-latex] (firstA.third east) -- (firstB.west);
\node[above = 1cm of firstB,label={[label distance=1mm]10:{\textbf{P1}}},inner sep=1pt]{};
}
};

\node[single,scale=0.3,right=3mm of first, inner sep=1mm] (second) {
  \tikz{
\node[quadruple] (secondA) at (-2,0) {$C^{\iota}_p(\overline{t}(\overline{a^{\iota}_g})(e^\zeta_g))=\langle\overline{C^{\alpha}_m}\rangle$
\nodepart{second}{$A^{\beta,\gamma}_g = \langle T(C^{\beta}_g,C^{\gamma}_g)\rangle$}
  \nodepart{third}{$\langle E^{\epsilon}_g\rangle^* = \overline{C^{\epsilon}_p}(a^{\beta,\gamma}_p)$}
  \nodepart{fourth}{$e^{\zeta}_g = \langle S[a^{\beta,\gamma}_g]({e^{\epsilon}}^*) \rangle$}
};
\node[single,right=6mm of secondA.four east] (secondB) {$\langle \overline{T}[\overline{A^{\iota}_g}](e^{\zeta}_g)\rangle$
};
\draw [-latex] (secondA.four east) -- (secondB.west);
\draw [-latex,dashed] (secondB.north) to[out=90,in=0] (secondA.one east);
\node[above = 1.5cm of secondB,label={[label distance=3mm]10:{\textbf{P2}}},inner sep=1pt]{};
}
};;

\draw[->] (first) -- (second);
\end{tikzpicture}
}
\end{minipage}

\caption{The current implementation \textbf{P1} prototypes automated evaluation
of blends according to user's arguments; this points to a proposed future implementation \textbf{P2}
with further automation.
 \label{fig:face-model}}
\end{figure}
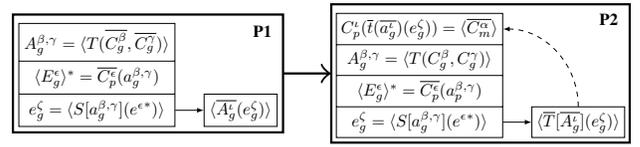

This is illustrated in Figure \ref{fig:face-model} in box \textbf{P1}, using the diagrammatic extension to the FACE model by \cite{colton-assessingprogress}.  Here, capital letters $F$, $A$, $C$, or $E$ are creative acts that generate a framing, aesthetic, concept, or example, respectively.  Administrative acts $S$ and $T$ denote selection and translation.  Lower-case letters denote the generated artefact in each case (e.g., the concept $c$ corresponding to the concept-creation act $C$).  Subscripts $p$, $g$, or $m$ indicate whether the act takes place at the process, ground, or meta level. Inside each box, stacks show the dependence in development epochs, and arrows show run-time message passing.  Acts taken by the programmer or user are decorated with a bar, whereas acts taken by the system itself receive no extra decoration.

In the current version of the system, apart from the \emph{programmer's} creative acts
specifying the modules and their interconnections, and the algorithm $\overline{C^{\epsilon}_p}$ that turns inputs into blends, the user, who is assumed to be a music expert,
must intervene in the system in two places.

\vspace{-.1cm}
First, the \emph{user} defines system settings $\overline{C^{\gamma}_g}$, $\overline{C^{\beta}_g}$  that correspond to the selection of  input idioms and of arguments respectively. Second, after the run completes, he or she evaluates the system output via $\overline{A^{\iota}_g}$.

The \emph{system's} primary responsibilities take place through the
creative acts $E^{\epsilon}_g$, which generate blends, and $S[a^{\beta,\gamma}_g]({e^{\epsilon}}^*)$, in
which the aesthetic $A^{\beta,\gamma}_g$ (a unified label for $\mathsf{assoc}$ and $\mathsf{asym}$, which are defined anew in each run, based on a fixed translation of the user's arguments, as specified in the previous section) is applied to rate the possible blends, and select to a final extended transition matrix.

Therefore, the key idea behind what has been implemented so far is an `automated
ranking/evaluation' step that guides the selection of blends,
$S[a^{\beta,\gamma}_g]({e^{\epsilon}}^*)$ according to the arguments defined by the user. The development of the programmatic components that operationalise this process has
relied on both computer science and musicological insights.
This approach has been characterised as meaningful per se through informal feedback provided by musical experts -- but is perhaps especially valuable because it
constitutes a prototype for more involved automated evaluation
of computer-generated harmonic spaces.

Indeed, the next step towards the development of a more autonomously
creative system using the same architecture is fairly clear: future
work would need to `close the loop' computationally, connecting the
evaluation of generated transition matrices with the parameter-setting (i.e., argumentation)
stage, and making this run autonomously to refine the system's
behavior.  This as-yet hypothetical situation is illustrated in the
box \textbf{P2}.

Here, the programmer has translated some of the user-specified aesthetics
into code $\overline{T}[\overline{A^{\iota}_g}]$, and invented a meta-level concept
$\overline{C^{\alpha}_m}$ defining a system component that can automatically apply 
these aesthetics to the generated transition matrices
$e^{\zeta}_g$ as in order to automatically
generate new system settings $C^{\gamma}_g$, $C^{\beta}_g$.  

\section{Conclusion, Discussion and Future Work}\label{sec:conclusion}

In this paper, we described a methodology for harmonic blending and we proposed a creative system that assists musicologists with the evaluation and enhancement of harmonic innovation. We defined some harmonic features of chord transitions utilised for evaluating blends of transitions, leading to the invention of novel harmonic spaces. The system allows a musicologist to specify arguments over these features that are taken into account in the generation of new harmonic spaces. The music expert can then assess whether the new harmonic idiom makes musicological sense and re-adjust the arguments to explore alternative blends that can potentially produce better harmonic spaces.

The main advantage of the current system is the agile interaction through which the user can express desirable properties over the transition blends and their argument-based evaluation in order to produce musically meaningful results. The added value of argumentation is the ranking/evaluation of blended transition -- obtained by conceptual blending of two input transition belonging to two musical idioms -- by answering questions which abstract several properties of chord transitions. On the other hand, the evaluation of the creative output of the system, i.e.,\ an extended harmonic space that includes blended transitions, is carried out by the user via an introspective argumentative dialogue.

In a future work we intent to use the argumentation-based process for evaluating the blended harmonisations of user defined melodies, i.e.,\ actual music output. Additionally, mapping the properties of the blended idiom or, at a latter stage of a harmonised melody, back to the parameter-setting stage opens an interesting direction for future research and further improvements of the system. The added value of argumentation can be stressed, for instance, by letting the system suggest possible refinements of the initial user arguments, progressively converting part of the introspective user evaluation into a more explicit format. For example, a future version of the system would be based on identifying harmonic features of the input spaces that automatically suggest an `optimal' set of initial arguments. The current version of the system is an already-usable prototype on the way towards the development of a more autonomous creative system.







\bibliographystyle{iccc}
\bibliography{iccc2016}

\end{document}